\documentclass[sigconf]{acmart}
\renewcommand\footnotetextcopyrightpermission[1]{} 
\usepackage[utf8]{inputenc} 
\usepackage{booktabs}       
\usepackage{amsfonts}       
\usepackage{nicefrac}       
\usepackage{microtype}      
\usepackage{xcolor}         
\usepackage[T1]{fontenc}
\usepackage{graphicx}
\usepackage[normalem]{ulem}
\usepackage{amsmath}

\usepackage{booktabs} 
\usepackage{tablefootnote} 
\usepackage{threeparttable} 
\usepackage{multirow}

\acmConference{JPMorgan Chase \& Co.}{2025}{Palo Alto,CA, USA}
  
\title{Better with Less: Small Proprietary Models Surpass Large Language Models in Financial Transaction Understanding}

\author{Wanying Ding}
\affiliation{
    \institution{JPMorgan Chase \& Co.}
    \city{Palo Alto}
    \state{California}
    \country{USA}
}
\email{wanying.ding@jpmchase.com}

\author{Savinay Narendra}
\affiliation{
    \institution{JPMorgan Chase \& Co.}
    \city{Palo Alto}
    \state{California}
    \country{USA}
}
\email{savinay.narendra@jpmchase.com}

\author{Xiran Shi}
\affiliation{
    \institution{JPMorgan Chase \& Co.}
    \city{Palo Alto}
    \state{California}
    \country{USA}
}
\email{xiran.shi@jpmchase.com}

\author{Adwait Ratnaparkhi}
\affiliation{
    \institution{JPMorgan Chase \& Co.}
    \city{Palo Alto}
    \state{California}
    \country{USA}
}
\email{adwait.ratnaparkhi@jpmchase.com}

\author{Chengrui Yang}
\affiliation{
    \institution{JPMorgan Chase \& Co.}
    \city{New York}
    \state{New York}
    \country{USA}
}
\email{chengrui.yang@chase.com}

\author{Nikoo Sabzevar}
\affiliation{
    \institution{JPMorgan Chase \& Co.}
    \city{Chicago}
    \state{Illinois}
    \country{USA}
}
\email{nikoo.sabzevar@chase.com}

\author{Ziyan Yin}
\affiliation{
    \institution{JPMorgan Chase \& Co.}
    \city{Wilmington}
    \state{Delaware}
    \country{USA}
}
\email{ziyan.yin@chase.com}

\begin{document}

\begin{abstract}
Analyzing financial transactions is crucial for ensuring regulatory compliance, detecting fraud, and supporting decisions. The complexity of financial transaction data necessitates advanced techniques to extract meaningful insights and ensure accurate analysis. Since Transformer-based models have shown outstanding performance across multiple domains, this paper seeks to explore their potential in understanding financial transactions. This paper conducts extensive experiments to evaluate three types of Transformer models: Encoder-Only, Decoder-Only, and Encoder-Decoder models. For each type, we explore three options: pretrained LLMs, fine-tuned LLMs, and small proprietary models developed from scratch. Our analysis reveals that while LLMs, such as LLaMA3-8b, Flan-T5, and SBERT, demonstrate impressive capabilities in various natural language processing tasks, they do not significantly outperform small proprietary models in the specific context of financial transaction understanding. This phenomenon is particularly evident in terms of speed and cost efficiency. Proprietary models, tailored to the unique requirements of transaction data, exhibit faster processing times and lower operational costs, making them more suitable for real-time applications in the financial sector. Our findings highlight the importance of model selection based on domain-specific needs and underscore the potential advantages of customized proprietary models over general-purpose LLMs in specialized applications. Ultimately, we chose to implement a proprietary decoder-only model to handle the complex transactions that we previously couldn't manage. This model can help us to improve 14\% transaction coverage, and save more than \$13 million annual cost.
\end{abstract}

\maketitle

\section{Introduction}

In the payments industry, the ability to accurately understand transactions is crucial for assessing business risks, enhancing user experiences, and minimizing inquiry costs. However, this task is complicated by the fact that transaction data is often messy and expressed in various formats. Traditional rule-based methods can manage data from a limited number of merchants, but they struggle to handle the vast array of unseen merchants, making them inefficient and unscalable. In contrast, the advanced language models, particularly pretrained large language models (LLMs) such as GPT\cite{radford2018improving}, Llama\cite{touvron2023llama}, T5\cite{raffel2020exploring}, and BERT\cite{kenton2019bert}, has demonstrated remarkable capabilities in analyzing complex data, offering a promising solution to the challenges. However, the application of LLMs in production environments comes with limitations. One significant drawback is the substantial computational resources required to run these models, which can lead to high operational costs and latency issues. Additionally, pretrained LLMs are not optimized for the unique requirements and nuances of particular applications, often necessitating extensive fine-tuning and optimization to perform well on specific tasks, which can be time-consuming and resource-intensive.

On the other hand, building a Transformer model from scratch can be necessary due to the unique and specialized nature for financial transaction data. Unlike general language data, transaction data often involves specific terminologies, unstructured formats, and domain-specific patterns that pretrained LLMs may not adequately capture. By developing a Transformer model from scratch, it is possible to design and train the model specifically for the intricacies of transaction data. This approach allows for the incorporation of domain-specific knowledge and the customization of the model architecture to better handle transactions. 

Choosing between pretrained LLMs and small proprietary Transformer models can be a complex and challenging task in real-world applications. This decision involves weighing various factors such as accuracy, scalability, cost, and the specific requirements of the application at hand. Our study makes a significant contribution by demonstrating that while LLMs excel in general tasks, a proprietary model is essential for specific tasks, such as financial transaction understanding. We provide a comprehensive comparison of pretrained LLMs and small proprietary Transformer models, evaluating both accuracy and efficiency. Our findings reveal that larger models do not always outperform smaller ones in practical settings. Additionally, we offer detailed guidance on choosing the right proprietary models for real-world applications, assisting decision-makers in navigating the complexities of model selection.

\section{Problem Definition}
\label{sc:problem}
\begin{figure}
\centering
\includegraphics[width=0.85\linewidth]{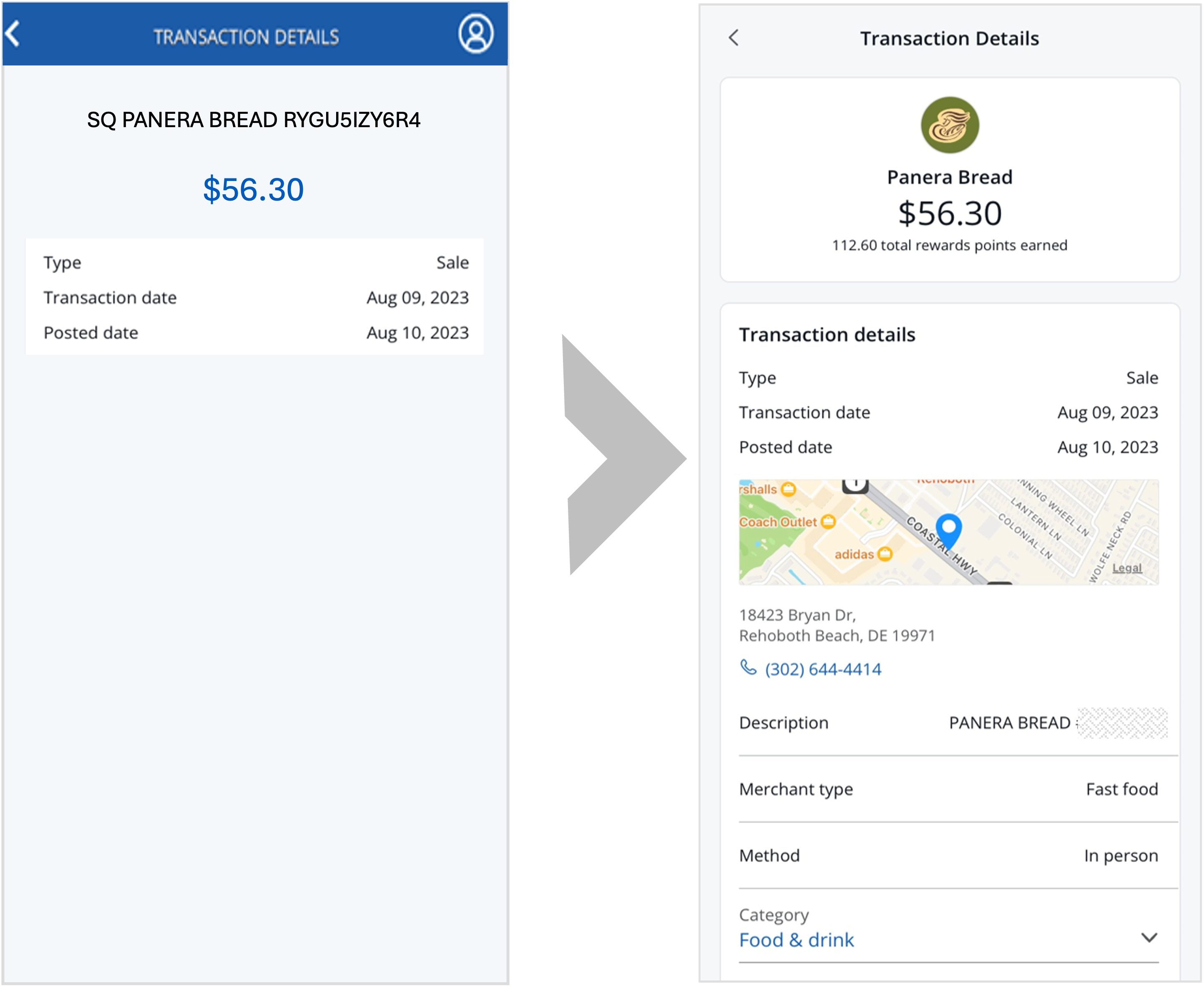}
\caption{Example for Transaction Understanding\label{fig:example}}
\end{figure}

The aim of this study is to standardize POS (point-of-sale) transactions with merchant information, which provide value-added features for Chase customers and offer merchant-level insights for the business. Our goal is to identify the corresponding merchant IDs from messy transaction text and enhance the merchant information available to users on our Chase mobile app (as shown in Figure \ref{fig:example}). Instead of displaying a raw transaction, we can present detailed merchant information, such as the logo, formal name, address, phone number, etc., to provide users with more context for inquiries and fraud detection.

\footnotetext{All transaction data mentioned in this paper is synthetic. Actual data is confidential}

The challenges stem from two main factors: the high volume and the highly disorganized nature of transactions. We need to process over 50 million transactions daily, requiring our model to respond within milliseconds while maintaining cost efficiency. Additionally, transactions are extremely complex and contain a significant amount of noise, making them difficult to manage with a simple solution. As illustrated in Table \ref{tb:txn_example}, numerous transactions lack an obvious sub-strings to identify merchants. For instance, "SWA * EARLYBRD XQQJWQ9V4F4" is difficult to associate with "Southwest Air". Furthermore, transactions often include various types of noise, such as aggregators (third-party online payment processors) like "SQ (Square)", but there is no consistent pattern for where and how these aggregators appear. Additionally, the presence of transaction numbers makes it more challenging. For instance, in "FAM EXP LDGX," the segment "LDGX" is a transaction number rather than part of the merchant's name. 
\begin{table*}[h]
\centering
\begin{tabular}{|l|l|l|l|}
\hline
Transaction. Text            & Transaction  Zipcode & Merchant ID & Merchant Name     \\ \hline
SQ * HM SP NTW P2FJOC4 & 12345                & JPMC001     & Home Shopping Network         \\ \hline
AUTOMA MSFT * CORPO008  & 67890                & JPMC002     & Microsoft    \\ \hline
SWA * EARLYBRD XQQJWQ9V4F4         & 13579                & JPMC003     & Southwest Air         \\ \hline
FAM EXP LDGX                & 24680                & JPMC004     & Family Express    \\ \hline
GOOGL * ADSJL0CZ1DN      & 98765                & JPMC005     & Google Ads \\ \hline
\end{tabular}
\par \begingroup\footnotesize\raggedright Data in the table is synthetic.\endgroup
\caption{Financial Transaction Examples}
\label{tb:txn_example}
\end{table*}

Our previous solution comprised two components: a set of regex rules (Rulebased) and an Enhanced String Distance (ESD) method. Rulebased method maintains thousands of regex rules to map incoming transactions to their corresponding merchant IDs. This method can only cover about 1000 merchants, which is highly unscalable and requires significant effort to create, modify, and maintain rules. The ESD method utilizes string similarity measures in conjunction with a manual decision tree, which is more flexible and can cover a larger number of merchants. However, the ESD method can only cover less than 20\% of transactions, leaving a significant amount of transactions unmanaged. These transactions lead to poor user experiences and numerous transaction inquiry calls, which annually cost the company millions of dollars. 

The purpose of this study is to explore advanced language models that can handle more complex transactions. Regarding advanced language models, Transformer-based models have emerged as the state-of-the-art (SOTA) in natural language processing (NLP). These models have demonstrated exceptional performance across a wide range of tasks due to their ability to capture intricate patterns and dependencies in data. In this study, we aim to explore both pretrained LLMs and small proprietary Transformer models. By comparing these approaches, we seek to determine the most effective solution for managing complex transactions, balancing performance, efficiency, and cost.

\section{Methodology}
\label{sc:method}
LLMs, pretrained on extensive public data that includes merchant nicknames and abbreviations, have significant potential to standardize transactions. Alternatively, training a proprietary model from scratch with specific transaction data also promises accurate predictions. To investigate both models, we have established six work streams centered around three types of Transformer-based models: Encoder-Only, Decoder-Only, and Encoder-Decoder Models. We will compare the pretrained versions of these models with their proprietary counterparts.


\subsection{Encoder Only Model}
When utilizing the Encoder-Only model to address this problem, we employ the Bi-Encoder framework\cite{reimers2019sentence}. In this method, we initially use an encoder model to convert all merchants in our database into embedding vectors, which are then stored in a vector database. When a transaction occurs, the same encoder model is used to encode the transaction. We then search for and rank the most likely merchants based on similarity, selecting the top-ranked merchant as the match with Lucene Vector Search\cite{xian2024vector}.

\subsubsection{SentenceBert}
The most widely used pretrained Encoder-Only model is SentenceBERT (SBERT) \cite{reimers2019sentence}. SBERT modifies the pretrained BERT network using Siamese and Triplet network\cite{schroff2015facenet}, enabling faster and more accurate semantic search by deriving semantically meaningful sentence embeddings that can be compared using cosine similarity. We utilized the all-MiniLM-L6-v2\cite{wang2020minilm} model from the MiniLM (Miniature Language Model) family. This model encodes both merchants and transactions into 384-dimensional dense vectors. During fine-tuning stage, we employed online contrastive loss to enhance the model's performance.  

\subsubsection{Proprietary Encoder-Only Model}
\label{sc:train_encoder_only}
We developed a proprietary encoder-only model using PyTorch by implementing the TransformerEncoderLayer module. This custom model is designed to encode both transactions and merchants with a unified encoder architecture. To train the model, we prepared a comprehensive dataset (see Section \ref{sc:dataset}) and ensured it was properly tokenized (see Section \ref{sc:tokenization}). We configured hyperparameters, such as learning rate and weight decay, with Ray Tune \cite{liaw2018tune} and Optuna \cite{akiba2019optuna}, but for other model structure hyperparamters, we setup a series experiments to find the best ones(see Section \ref{sc:experiments}). The model was trained using a contrastive loss function, as specified in Equation \ref{eq:encoder_loss}.
\begin{equation}
L = (1-\text{pos\_sim}) + \max(0, \text{neg\_sim} - \text{margin})
\label{eq:encoder_loss}
\end{equation}
where the $pos\_sim$ indicates the cosine similarity between transaction and the positive merchant name, and $neg\_sim$ indicates the cosine similarity between transaction and the negative merchant name. We set margin as 0.5 on our data to achieve optimal results.  

\subsection{Decoder Only Model}
When using the Decoder-Only model, we treat this transaction understanding problem as a translation problem, which generates a standardized merchant name from the messy transaction text. With the generated merchant name, we search our merchant database to find the most possible merchant ID with Lucene String Search.
\subsubsection{Llama3-8b}
Transaction data is confidential, so we only utilize open-source models that can be deployed internally, which is why we opted for Llama3 instead of GPT-4. Llama3\cite{touvron2023llama} is the latest foundational LLM from Meta and achieves comparable results to GPT-4 in various benchmarks. Since the full version of Llama3 is unnecessary, we implement the Llama3-8b, which is optimized for dialogue use cases. Here is the given prompt to Llama3-8b: 
\begin{quote}
You are provided with raw merchant transaction text and the transaction's zipcode. Based on the provided information, please output the merchant name. Your answer should only include the predicted merchant name and nothing else.
\end{quote}followed with a user message with a transaction text and zipcode. 

During the fine-tuning stage, we also include the corresponding merchant name in the assistant message. We applied QLoRA\cite{dettmers2024qlora} to fine-tune the model.

\subsubsection{Proprietary Decoder-Only Model}
\label{sc:train_decoder_only}
The proprietary Decoder-Only model is built using PyTorch's TransformerDecoderLayer module. In our implementation, the model takes transactions text as input and decodes them to desired merchant names as output. The model is trained in the similar way as described in Proprietary Encoder-Only Model part(Section \ref{sc:train_encoder_only}). We applied Cross Entropy (Equation\ref{eq:decoder_loss}) as the loss function to train the model. 

\begin{equation}
L = -\frac{1}{N} \sum_{i=1}^{N} \sum_{c=1}^{C} y_{i,c} \log(\hat{y}_{i,c})
\label{eq:decoder_loss}
\end{equation}
where $N$ is the number of tokens in the sequence, $C$ is the vocabulary size.

\subsection{Encoder-Decoder Model}
The Encoder-Decoder model also translates transactions to merchant names, the same as Decoder-Only model does. The primary distinction between the Decoder-Only model and the Encoder-Decoder model lies in the additional encoding step present in the latter. Specifically, the Encoder-Decoder model employs several layers of encoders to transform a transaction into a context representation. The decoder then uses this representation, instead of transaction tokens, to generate output.
\subsubsection{Flan-T5}
Flan-T5\cite{chung2024scaling} is a widely used pretrained Encoder-Decoder model, which is a family of models that enhances T5\cite{raffel2020exploring}. While fine-tuning, we follow the same settings to Llama3-8b. 
\subsubsection{Proprietary Encoder-Decoder Model}
Similarly, we apply PyTorch's TransfromerLayer to build up our proprietary Encoder-Decoder model. The transaction data will first go through several layers of TransformerEncoderLayer and then leverage several layers of TransformerDecoderLayer to decode out the merchant name. The model training configuration and loss function is the same with Decoder-Only model (Section\ref{sc:train_decoder_only}). 

\section{Experiments}
\label{sc:dataset}
\subsection{Transaction Dataset}
We categorize our transaction data into three types: Rulebased Data, which consists of transactions managed by existing regex rules; ESD Data, which includes transactions handled by Enhanced String Distance (ESD) method; and Rawcleansed Data, which refers to transactions that are not managed by any existing methods and stored in their original form as they are received. Rulebased data takes about 63\% of our dataset, ESD takes about 17\%, and Rawcleansed takes about 20\%. 
\begin{itemize}
\item Rulebased Data: 
We conducted a query of our historical transactions and extracted a sample of approximately 1,131 rules from our dataset. For the training dataset, we compiled 773,653 transactions representing 779 merchants. Additionally, we prepared a testing dataset consisting of 1,311 transactions, also covering these 779 merchants. 
\item {ESD Data}:
We queried our historical transactions and sampled 574,871 transactions managed by the ESD method, encompassing 506,135 merchants. For the testing phase, we prepared two distinct ESD test datasets. The first dataset, ESD\_RD, includes 40,223 transactions involving merchants present in the training data. The second dataset, ESD\_ZS, comprises 10,000 transactions involving merchants not previously seen in the training data.
\item {Rawcleansed}:
In the absence of ground truth for the Rawcleansed data, we manually labeled the merchants for a set of 2,541 transactions. This labeled dataset now serves as the ground truth for evaluating the model's performance on Rawcleansed data. 
\end{itemize}

To train the Encoder-Only model effectively, the inclusion of negative samples is essential (shown as in Equation\ref{eq:encoder_loss}). For each transaction, the corresponding matched merchant serves as the positive sample. We utilize Jaccard Similarity to compare the positive merchant with all other merchants. A merchant with a similarity score greater than 0.75 but less than 1.0 is randomly chosen as the negative sample. This approach ensures that the encoder model can differentiate between merchants that are similar but not identical.

\subsection{Merchant Dataset}
We compiled a dataset of 7.8 million merchants for matching purposes. To refine the candidate pool, we use the merchant's and transaction's zipcode as a filter. Following this, we apply either the generated merchant names or the encoded transaction vectors to the filtered candidates for more precise matching via Lucene, and select the top 1 merchant as the predicted match.

\subsection{Evaluation Metrics}
In our task, we use `Accuracy` to evaluate models' performance. The Accuracy means the ratio of transactions that can be correctly map to a right merchant. Since we have four datasets, we apply a `Weighted Accuracy` to evaluate each model's performance. The Weighted Accuracy is calculated as 
\begin{align*}
    \text{Weighted Accuracy}= &0.63*\text{Rulebased Accuracy}\\
                            &+0.085*\text{ESD\_RD Accuracy} \\
                            &+0.085*\text{ESD\_ZS Accuracy}\\
                            &+0.2*\text{Rawcleansed Accuracy}
\end{align*}
where the weight for each part is derived from the percentage of this data present in our dataset.

\section{Results and Discussions}
\label{sc:experiments}
\subsection{Open Sourced or Proprietary  Model}
\begin{figure*}
\centering
\includegraphics[width=0.85\linewidth]{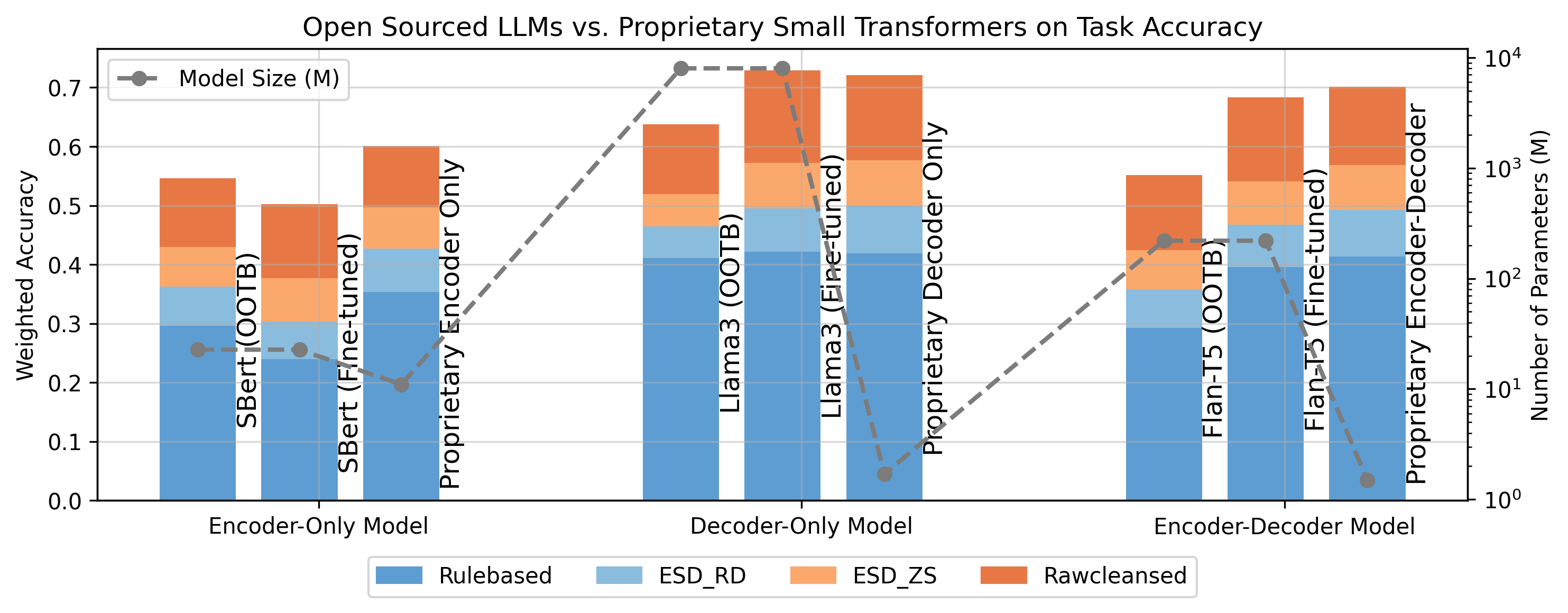}
\caption{Open Sourced LLMs vs. Proprietary Small Transformers on Task Accuracy}
\label{fig:open_vs_house}
\end{figure*}

In this section, we evaluated three model streams: the out-of-the-box (OOTB) LLM, the fine-tuned LLM, and the proprietary model. To cover a broad range of LLMs, we selected Llama3 (decoder-only model), Flan-T5 (encoder-decoder model), and SBert (encoder-only model) as our LLM candidates. Correspondingly, we developed three types of proprietary models: a Proprietary Decoder-Only model, a Proprietary Encoder-Decoder model, and a Proprietary Encoder-Only model for comparison (shown as Figure\ref{fig:open_vs_house}). The LLMs were fine-tuned with the exact same datasets used to train the proprietary models. 

As shown in Figure \ref{fig:open_vs_house}, OOTB models are not well-suited for our internal uses cases. Although these models are designed for general purposes, they do not perform effectively on specific business scenarios because they lack exposure to the target data. After fine-tuning, their performance improves significantly since fine-tuning allows the models to adapt specifically to the nuances and patterns of the target data. Llama3's performance increases 9\% and Flan-T5's performance increases 13\% after fine-tuning. However, SBert's performance decreases 4\%. The possible reason is that the transaction data is not well aligned with the model structure and original dataset so the transaction data introduces noises and biases that the model was not previously exposed to. 

However, upon examining the proprietary models, we find that they can actually perform comparably or even better than LLMs with significantly fewer parameters. For instance, the Proprietary Decoder-Only model achieves 72.07\% accuracy with just 1.7M parameters, compared to Llama3's 72.89\% accuracy with 8B parameters. Proprietary Encoder-Decoder model performs better than Flan-T5 by 2\% (70.14\% vs. 68.31\%) with much less parameters (1.5M vs 220M), and proprietary Encoder-Only model also performances better than SBert by 6\% (60.06\% vs. 54.63\%) with smaller size (11M vs 22.7M). This suggests that proprietary models can be more efficient in certain tasks compared to larger pretrained models. 
Table \ref{tb:open_vs_house_cost} \footnote{The price is estimated according to the AWS pricing page: https://aws.amazon.com/sagemaker-ai/pricing/} provides detailed insights into the speed and cost of training/fine-tuning each of these models. It is evident that proprietary models are smaller and faster in both training and inference, resulting in budget savings. Given that we need to process over 50+ million transactions per day, a 100-millisecond upper limit for inference time is crucial. Although Llama3 achieves 0.8\% higher accuracy, it requires more than 700 milliseconds to infer one transaction. In contrast, the Proprietary Decoder-Only model takes only 95.02 milliseconds to achieve comparable accuracy, offering much faster inference time and reduced costs. Proprietary Decoder-Only model achieves better accuracy compared to the Proprietary Encoder-Decoder model (72.07\% vs. 70.14\%), but it is significantly slower in inference despite having similar model sizes. The possible reason is that the encoder-decoder model architecture can efficiently utilize parallel processing during the encoding phase and make use of the encoded context for effective decoding. In contrast, the decoder-only model architecture has to access a rich context from the start, which increases the complexity of each decoding step. Additionally, in a decoder-only model, the entire process of understanding the input and generating the output is handled in a single pass. This requires the model to rely on previously generated tokens to predict the next one, known as autoregressive generation, which can slow down the overall process. Encoder models are faster and cheaper, even with much larger model size. This is because the encoding process can be well paralleled. However, their performance on this task is not as good as models with decoders.

\begin{table*}[]
\begin{tabular}{|c|c|c|c|c|c|c|}
\hline
Model                                                                 & Model Size & Machine                                                                                    & Training Speed & Training Cost    & Inference Time                                                          & Inference Cost  \\ \hline
Llama 3                                                               & 8B         & \begin{tabular}[c]{@{}c@{}}Training : g4dn.12xlarge\\ Inference: g4dn.2xlarge\end{tabular} & 166h/1M txn    & \$ 812/1M txn    & \begin{tabular}[c]{@{}c@{}}735 ms/txn \\ 204.17h/1M txn\end{tabular}    & \$191/1M txn    \\ \hline
Flan-T5                                                               & 220M       & \begin{tabular}[c]{@{}c@{}}Training : g4dn.12xlarge\\ Inference: g4dn.2xlarge\end{tabular} & 108h/1M txn    & \$528/1M txn     & \begin{tabular}[c]{@{}c@{}}284.7 ms/txn \\ 3.52h/1M   txn\end{tabular}  & \$22.56/1M txn  \\ \hline
SBert                                                                 & 22.7M       & g4dn.xlarge                                                                               & 1.7h/1M   txn   & \$1.60/1M   txn & \begin{tabular}[c]{@{}c@{}}5.77   ms/txn \\ 1.60 h/1M txn\end{tabular} & \$1.51/1M   txn \\ \hline
\begin{tabular}[c]{@{}c@{}}Proprietary\\ Decoder Only\end{tabular}    & 1.7M       & g4dn.xlarge                                                                                & 0.33h/1M txn   & \$0.24/1M txn    & \begin{tabular}[c]{@{}c@{}}95.02 ms/txn \\ 26.39h/1M   txn\end{tabular} & \$19.43/1M txn  \\ \hline
\begin{tabular}[c]{@{}c@{}}Proprietary\\ Encoder-Decoder\end{tabular} & 1.5 M      & g4dn.xlarge                                                                                & 0.11h/1M txn   & \$0.08/1M txn    & \begin{tabular}[c]{@{}c@{}}26.50 ms/txn\\  7.36h/1M txn\end{tabular}    & \$5.42/1M txn   \\ \hline
\begin{tabular}[c]{@{}c@{}}Proprietary\\  Encoder Only\end{tabular}   & 11.3M        & g4dn.xlarge                                                                                & 0.46h/1M txn   & \$0.34/1M txn    & \begin{tabular}[c]{@{}c@{}}7.62 ms/txn \\  2.11h/1M   txn\end{tabular} & \$1.57/1M txn   \\ \hline

\end{tabular}
\caption{Open Sourced LLMs vs. Proprietary Small Transformers on Speed and Cost}
\label{tb:open_vs_house_cost}
\end{table*}

\subsection{The Larger the Better?}
From the results in Figure \ref{fig:open_vs_house}, we are prompted to ask whether a larger model necessarily leads to better outcomes. We use a series of experiments to explore this question. Speaking of model size, three factors contribute to it, the vocabulary size, and number of layers, and embedding size. So, we discuss from these three dimensions.

\subsubsection{Tokenizer and Vocabulary Size}
\label{sc:tokenization}
A tokenizer is a key part of Transformer-based models, essential for preparing text data. It converts raw text into a format the model can process by breaking it down into smaller units called tokens, which can be words, subwords, or characters. Common tokenizers include BPE, WordPiece, and Unigram. BPE\cite{sennrich2015neural} starts with individual characters, and merge the most frequent pairs of characters, until pre-defined vocabulary size is reached. WordPiece\cite{wu2016google} starts with individual characters and adds the most frequent subword units to the vocabulary to maximize the training data's likelihood. Unigram\cite{kudo2018subword} uses a probabilistic model treating subword units as independent tokens. It starts with a large vocabulary and removes the least likely units to maximize the training data's likelihood. Given that there is no definitive answer regarding the best tokenizer or the optimal vocabulary size, we address this issue by experimenting with all three popular tokenizers mentioned above, and vary the vocabulary size from 100 to 10,000 to check whether a larger vocabulary yields better results.

\begin{figure*}
\centering
\includegraphics[width=0.9\linewidth]{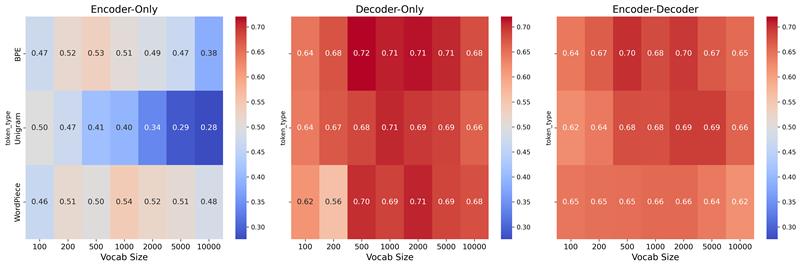}
\par The number showing in the heatmap is the Weighted Accuracy with the corresponding tokenizer and vocabulary size. To make a fair comparison, all the models are using similar configurations with 8 layers and 128 embedding size. 
\caption{Models' performance on with different tokenizers and vocabulary size}
\label{fig:vocab_size}
\end{figure*}

Figure \ref{fig:vocab_size} addresses both questions, emphasizing the importance of selecting the right tokenizer and vocabulary size for optimal performance. The Encoder-Only model performs best with WordPiece and significantly worse with Unigram. On the other hand, the Encoder-Decoder model yields better results with BPE compared to WordPiece. The Decoder-Only model seems less affected by the choice of tokenizer, performing well with all three options. 

The Encoder-Only model benefits from WordPiece because WordPiece is designed to create a balance between vocabulary size and token granularity. It segments words into subword units that are frequent enough to capture meaningful linguistic information. In contrast, Unigram uses a probabilistic approach to select subword units, which sometimes leads to less optimal segmentation and produce a larger variety of subword units that are not meaningful. BPE is designed to iteratively merge the most frequent pairs of bytes, resulting in a compact set of subword units. This compression is beneficial for the Encoder-Decoder models, which need to handle both input and output sequences efficiently, especially when transforming sequences of similar lengths. The Decoder-Only model's insensitivity to tokenizer choice suggests its architecture is robust enough to handle various tokenization strategies without significant performance changes.

Regarding vocabulary size, bigger is not always better. The Encoder-Only model achieves its best performance with a vocabulary size of 1000, which might provide the right balance between having enough subword units to capture nuances and not overwhelming the model with too many options, which can complicate learning. While both the Decoder-Only and Encoder-Decoder models perform optimally with a smaller vocabulary size of 500, because it reduces the complexity of the output space, making it easier for the model to learn and generate sequences effectively. 

In our subsequent experiments, we use the WordPiece tokenizer with a vocabulary size of 1000 for the Encoder-Only model and BPE tokenizers with a vocabulary size of 500 for both the Encoder-Decoder model and Decoder-Only model.
\subsubsection{Model Size}
\begin{figure*}
\centering
\includegraphics[width=0.85\linewidth]{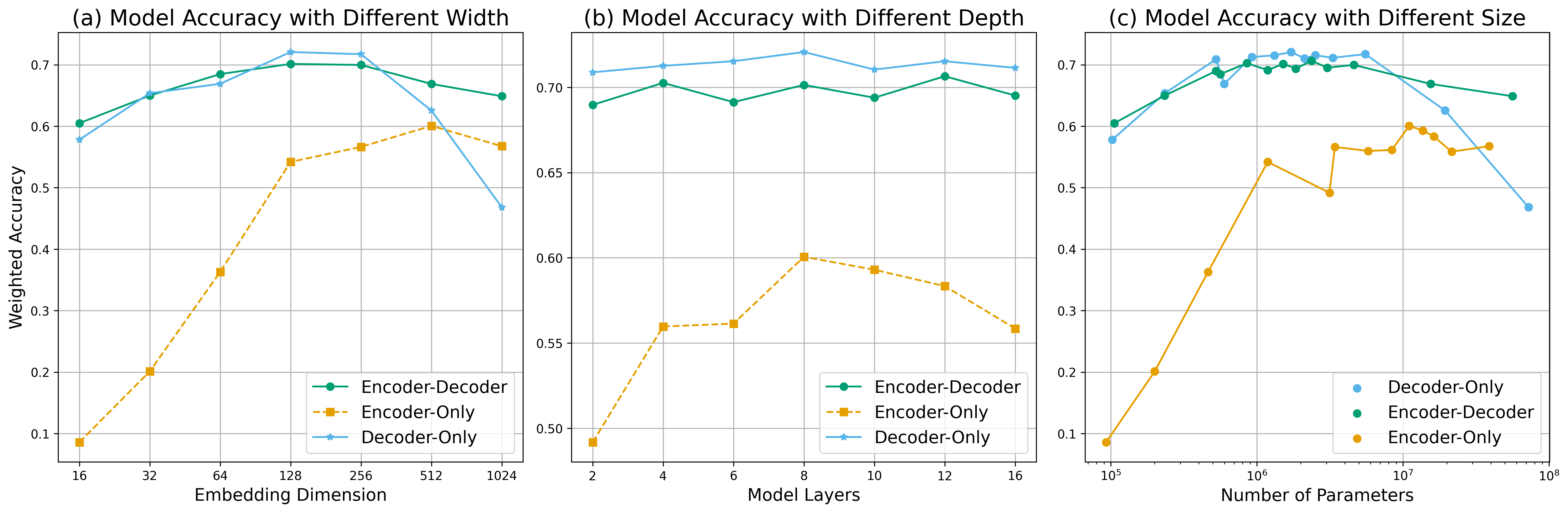}
\caption{Models' performances with different Model Size}
\label{fig:model_size}
\end{figure*}

After setting up the vocabulary and tokenizer, we will discuss about whether a model with more parameters will yield better results. We first vary the embedding dimension for the three models from 16 to 1024 while keeping the model depth fixed at 8. As shown in Figure \ref{fig:model_size}(a), increasing the embedding size generally enhances model performance. However, the Encoder-Only model is particularly sensitive to this parameter, as it requires a wide embedding size to effectively store context information. The Encoder-Only model achieves its best performance at an embedding size of 512, after which performance declines with larger sizes. Similarly, the Decoder-Only model's performance improves initially but drops significantly when the embedding size exceeds 256. The Encoder-Decoder model, on the other hand, demonstrates stability across different embedding sizes, although its accuracy slightly increases and then begins to decline after reaching 128.

We vary the number of layers from 2 to 16 for the three models. In the Encoder-Decoder model, layers are evenly split, so an 8-layer Encoder-Decoder model has 4 encoder layers and 4 decoder layers. As shown in Figure \ref{fig:model_size}(b), the Encoder-Only model is particularly sensitive to this parameter, with accuracy increasing dramatically at first, peaking at 8 layers, and then declining. In contrast, the Decoder-Only and Encoder-Decoder models exhibit greater stability with respect to the number of layers, indicating that model depth has less impact on their performance.

Figure \ref{fig:model_size}(c) provides a comprehensive overview of the models' performances relative to their sizes, which are determined by either the model's width (embedding size) or depth (number of layers). The results indicate that the Encoder-Only model benefits from a larger model size, achieving better performance with increased width or depth. In contrast, the Decoder-Only model performs better with small-to-medium model sizes, as larger sizes tend to introduce noise rather than useful information. The Encoder-Decoder model demonstrates greater stability across different model sizes.

\subsubsection{Model Type}
\begin{table*}[h!]
\centering
\begin{tabular}{|c|c|c|c|c|c|c|c|c|c|c|c|c|c|c|}
\hline
\multirow{2}{*}{Model Type} & 
\multirow{2}{*}{Tokenizer} & 
\multirow{2}{*}{V} & 
\multirow{2}{*}{D} & 
\multirow{2}{*}{L} & 
\multirow{2}{*}{N} & 
\multirow{2}{*}{\begin{tabular}[c]{@{}c@{}}Train\\Iters\end{tabular}} & 
\multirow{2}{*}{\begin{tabular}[c]{@{}c@{}}Train\\Time\end{tabular}} & 
\multirow{2}{*}{\begin{tabular}[c]{@{}c@{}}Index\\Time\end{tabular}} & 
\multirow{2}{*}{\begin{tabular}[c]{@{}c@{}}Infer\\Time (/txn)\end{tabular}} & 
\multicolumn{5}{c|}{Accuracy} \\
\cline{11-15}
 & & & & & & & & & & Top-1 & Top-5 & Precision & Recall & F1 \\
\hline
                              &                            &                    &                    &                    &                    &                                                                              &                                                                           &                                                                           &                                                                                   & \multicolumn{1}{c|}{Rule} & \multicolumn{1}{c|}{ERD}  & \multicolumn{1}{c|}{EZS}  & \multicolumn{1}{c|}{Raw}  & WT \\ \hline
Decoder Only                  & BPE                        & 500                & 128                & 8                  & 1.72M            & 10499                                                                        & 2.3 hrs                                                               & 6.5 mins                                                                  & 95.02 ms                                                                          & \multicolumn{1}{c|}{0.66} & \multicolumn{1}{c|}{0.95} & \multicolumn{1}{c|}{0.91} & \multicolumn{1}{c|}{0.72} & 0.72     \\ \hline
Encoder Decoder               & BPE                        & 500                & 128                & 8                  & 1.52M            & 4599                                                                         & 1.3 hrs                                                                & 6.5 mins                                                                  & 26.50 ms                                                                          & \multicolumn{1}{c|}{0.66} & \multicolumn{1}{c|}{0.92} & \multicolumn{1}{c|}{0.90} & \multicolumn{1}{c|}{0.66} & 0.70     \\ \hline
Encoder Only                  & WordPiece                  & 1000               & 512                & 8                  & 11.03M         & 899                                                                          & 2.2 hrs                                                               & 80 hrs                                                                    & 11.62 ms                                                                          & \multicolumn{1}{c|}{0.56} & \multicolumn{1}{c|}{0.87} & \multicolumn{1}{c|}{0.82} & \multicolumn{1}{c|}{0.52} & 0.60     \\ \hline
\end{tabular}
\par V indicates the vocabulary size, D indicates the embedding dimension, L indicates the number of layers, and N indicates the number of parameters. Rule is the Accuracy on Rulebased dataset, ERD is the Accuracy on ESD\_RD dataset, EZS is the Accuracy on ESD\_ZS dataset, and Raw is the Accuracy on Rawcleansed dataset, and WT is the final Weighted Accuracy. All these models are trained with the same batch size and similar learning rate.
\caption{Comparison Among Proprietary Models}
\label{tb:in_house_models}
\end{table*}

We have tried three different types of Transformer model in our task, namely Encoder-Only, Decoder-Only, and Encoder-Decoder model. As shown in Figure\ref{fig:vocab_size} and Figure\ref{fig:model_size}, Decoder-Only and Encoder-Decoder models outperform Encoder-Only models. Decoder-Only model and Encoder-Decoder model benefit from their autoregressive nature in the decoder part, which allows them to generate tokens sequentially and maintain coherence, making them particularly effective for tasks like text generation. When comparing the Encoder-Decoder model to the Decoder-Only model, the Decoder-Only model slightly outperforms the Encoder-Decoder model. This performance difference may be attributed to the potential information loss that occurs during the encoding process, leading to less accurate decoding compared to the Decoder-Only model, which does not undergo this encoding step.

Table \ref{tb:in_house_models} presents a comparison among the three types of proprietary models. For a fair evaluation, we selected the best-performing model from each category. The Encoder-Only model converges quickly, requiring only 899 iterations, while the Decoder-Only model takes longer to train, converging after 10,499 iterations. Despite the Encoder-Only model being larger, both models take a similar amount of time to converge. During inference, the Encoder-Only model is significantly faster, even with its larger number of parameters. However, the Encoder-Only model requires a longer time for the vector level indexing, and the index must be rebuilt each time the model is updated. In contrast, the Decoder-Only and Encoder-Decoder model are quicker to index and does not require re-indexing upon model updates, as the indexing is performed at the merchant level. Although both models are slower during inference, they still satisfy our 100ms/txn requirement.

If we investigate the Accuracy on each type of dataset, all models perform comparably better on ESD data compared to Rulebased data and Rawcleansed data. This indicates that the model effectively leverages similar patterns within the data for matching purposes. The performance on Rulebased transactions falls significantly,which is likely due to that Rulebased data contains business logic that could be captured by string-level patterns. The Rawcleansed data is unseen in our training data and of higher complexity, but Decoder-Only model can still get 72\% accuracy, dramatically higher than the other two models. This highlights the potential to utilize the Decoder-Only model to improve the coverage of our transaction system. 

\section{Model Deployment}
\begin{figure*}
\centering
\includegraphics[width=0.95\linewidth]{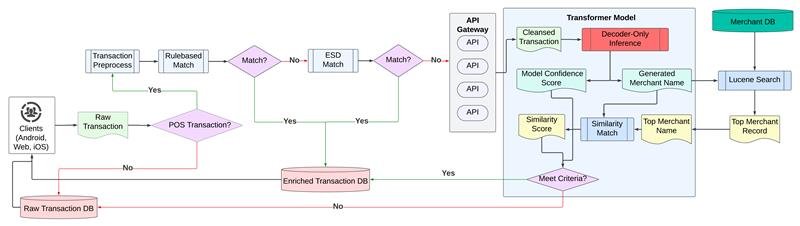}
\caption{Model Deployment}
\label{fig:deployment}
\end{figure*}

Since the Decoder-Only model achieves 72\% accuracy on Rawcleansed data, and our business priority is expanding coverage on Rawcleansed data. In the initial deployment, we use the Proprietary Decoder-Only model for Rawcleansed data data while retaining the Rulebased and ESD components for further enhancement.

The deployment pipeline is illustrated in Figure \ref{fig:deployment}. For POS transactions, we first apply general string cleaning, including lowercase conversion and non-alphanumeric removal. The transaction then passes through our Rulebased matching and ESD matching components. Transactions that do not find a match, known as Rawcleansed transactions, are processed by the Proprietary Decoder-Only model via API calls. The model outputs a confidence score and a generated merchant name for each transaction. We then use the generated merchant name for a Lucene search to identify the top merchant. However, this top merchant is only correct 72\% of the time, and we aim to avoid showing incorrect predictions to users. Therefore, we use the similarity between the top merchant's name and the generated merchant name, along with the model's confidence score, as filters to eliminate incorrect predictions. Monthly, we manually sample and review results, gathering incorrectly matched and inquired transactions to fine-tune the model.
\begin{table}
\begin{tabular}{|c|c|c|c|c|c|}
\hline
\multirow{2}{*}{Your First Column} & \multicolumn{4}{c|}{Transaction Coverage} & 
\multirow{2}{*}{\begin{tabular}[c]{@{}c@{}}Cost \\ Reduction\end{tabular}} \\ 
\cline{2-5}
 & \multicolumn{1}{l|}{Rule} & \multicolumn{1}{l|}{ESD} & \multicolumn{1}{l|}{Raw} & \multicolumn{1}{l|}{All} & \\ 
\hline
Before            & \multicolumn{1}{c|}{63\%} & \multicolumn{1}{c|}{17\%} & \multicolumn{1}{c|}{0\%}   & 80\%                     & --                                                                         \\ \hline
After             & \multicolumn{1}{c|}{63\%} & \multicolumn{1}{c|}{17\%} & \multicolumn{1}{c|}{14\%} & 94\%                     & \$13.2M/year                                                               \\ \hline
\end{tabular}
\caption{Business Metrics After Decoder Model Deployment}
\label{tb:deployment}
\end{table}

Table \ref{tb:deployment} provides an overview of the business metrics following the deployment of the Decoder-Only model. Currently, we retained the Rulebased and ESD components while focusing solely on the Rawcleansed data, which our previous system could not process. Rawcleansed data accounts for 20\% of our transactions, and the Decoder-Only model accurately predicts 72\% of these transactions. This leads to a 14\% increase in processed transactions, raising our overall transaction coverage from 80\% to 94\%. As a result, the improved coverage helps reduce transaction inquiry calls, generating an estimated annual savings of approximately \$13.2 million.

\section{Related Work}
Transformer models have revolutionized the field of natural language processing (NLP) since their introduction\cite{vaswani2017attention}. These models use self-attention to capture sequence dependencies, enabling efficient and accurate language processing. They fall into three types: Encoder-Only, Decoder-Only, and Encoder-Decoder models..\textbf{Encoder-Only Models} such as BERT\cite{kenton2019bert}, focus on understanding the input sequence by capturing contextual information from both directions. A notable example of an encoder-only model is SBERT\cite{reimers2019sentence}, which modifies BERT to generate semantically meaningful sentence representations. \textbf{Decoder-Only Models} such as GPT\cite{radford2019language,brown2020language}, are designed for text generation tasks. These models generate text by predicting the next token in a sequence, leveraging autoregressive mechanisms. Decoder-only models excel in tasks like text completion and dialogue generation.\textbf{Encoder-Decoder Model}  such as T5 (Text-To-Text Transfer Transformer), combine the strengths of both encoders and decoders. The encoder processes the input sequence to generate a context representation, which the decoder then uses to generate the output sequence. This architecture is particularly effective for tasks like machine translation, summarization, and question answering.

Large Language Models (LLMs) are advanced AI systems with vast parameters and training data, enabling them to capture complex linguistic patterns. Built on Transformer architecture, they excel in diverse NLP tasks across various fields. LLM integration into finance also becomes a significant area of research and exploration. Yang et al. introduced FinBERT\cite{yang2020finbert}, a BERT-based model fine-tuned for financial analysis, demonstrating its effectiveness in capturing the nuances of finance market. Bloomberg released BloombergGPT\cite{wu2023bloomberggpt}, a LLM specifically designed to enhance financial NLP tasks. Additionally, Xiao-Yang Liu et al. presented FinGPT\cite{yang2023fingpt}, aimed at enhancing tasks such as market prediction and financial data interpretation. However, much of this research has focused on developing general financial LLMs or applying them to general tasks like sentiment analysis or named entity resolution. Few studies have concentrated on specific tasks such as Financial Transaction Understanding. The Slope company\footnote{https://slopepay.com/company} has proposed two LLM based transaction understanding models: SlopeGPT\footnote{https://medium.com/slope-stories/slopegpt-the-first-payments-risk-model-powered-by-gpt-4-cd444ab5242d} and SlopeTransFormer\footnote{https://medium.com/slope-stories/slope-transformer-the-first-llm-trained-to-understand-the-language-of-banks-88adbb6c8da9}, but both models are fine-tuned from existing LLMs, and they do not provide detailed accuracy comparisons or explore the potential of using smaller models to achieve better performance.

\section{Conclusion}
This paper presents a comprehensive comparison between LLMs and small proprietary models within the domain of financial transaction understanding. We conducted extensive experiments on three types of Transformer models: Encoder-Only, Decoder-Only, and Encoder-Decoder, using both LLMs and small proprietary models. Our results indicate that small proprietary models offer comparable accuracy, faster processing and lower costs, making them more suitable for real-time financial applications. These findings emphasize the importance of selecting models based on domain-specific needs and highlight the potential benefits of customized proprietary models in specialized settings. In production, we chose to deploy the Proprietary Decoder-Only model to handle previously unmanageable
complex transactions. This improved transaction coverage by 14\% and reduced annual costs by over \$13 million.

\bibliographystyle{ACM-Reference-Format}
\bibliography{reference}

\end{document}